\documentstyle[12pt]{article}
\input amssym.def
\input amssym.tex

\newcommand{\wix}{\widetilde{x}}
\newcommand{\wip}{\widetilde{p}}

\begin{document}

\renewcommand{\theequation}{\thesection.\arabic{equation}}

\begin{titlepage}{\LARGE
\begin{center} A discrete time peakons lattice \end{center}}

\vspace{1.5cm}

\begin{flushleft}{\large Yuri B. SURIS}\end{flushleft} \vspace{1.0cm}
Centre for Complex Systems and Visualization, Unversity of Bremen,\\
Postfach 330 440, 28334 Bremen, Germany\\
e-mail: suris @ mathematik.uni-bremen.de

\vspace{2.0cm}

{\small {\bf Abstract.} A discretization of the peakons lattice is introduced,
belonging to the same hierarchy as the continuous--time system. The
construction examplifies the general scheme for integrable discretization
of systems on Lie algebras with $r$--matrix Poisson brackets.
An initial value problem for the difference equations is solved in terms of a
factorization problem in a group. Interpolating Hamiltonian flow is found.
A variational (Lagrangian) formulation is also given.}
\end{titlepage}

\setcounter{equation}{0}
\section{Introduction}
The subject of integrable symplectic maps received in the recent
years a considerable attention.
Given an integrable system of ordinary differential equations with such
attributes as Lax pair, $r$--matrix and so on, one would like to construct
its difference approximation, desirably also with a (discrete--time analog of)
Lax pair, $r$--matrix etc. Recent years brought us several successful
examples of such a construction [1--9].

Recently there appeared for the first time examples \cite{Discr CM},
\cite{Discr RS} where the Lax matrix of the
discrete--time approximation {\it coincides} with the Lax matrix of the
continuous--time system, so that the discrete--time system belongs to the
{\it same} integrable hierarchy as the underlying continuous--time one
(systems of Calogero--Moser type). This led us to the formulation of a
general recipe for producing  discretizations of this type, which was applied
to the two favorite examples of finite dimensional integrable systems of the
classical mechanics -- the Toda lattice and the relativistic Toda lattice
\cite{New discr Toda}, \cite{Discr RTL}.

In the present Letter we want to describe an application of this scheme to
another remarkable system, discovered recently by Camassa and Holm \cite{CH}.
Such an application was made possible thanks to an $r$--matrix interpretation
given to this system by Ragnisco and Bruschi \cite{RB}.

\setcounter{equation}{0}
\section{Continuous--time peakons lattice}
The peakons lattice has appeared in \cite{CH} as a system describing positions
and velocities of ''peaked solitons'' for a certain integrable shallow water
equation. The Hamiltonian function of the peakons lattice reads:
\[
H(x,p)=\frac{1}{2}\sum_{j,k=1}^Np_jp_k\exp(-|x_k-x_j|).
\]
It can be proved \cite{CH} that, generally speaking, the peakons retain their
ordering, so that one can assume $x_1>x_2>\ldots>x_N$, and the Hamiltonian
function takes the form
\begin{equation}\label{Ham}
H(x,p)=\frac{1}{2}\sum_{k=1}^Np_k^2+\sum_{1\le j<k\le N}p_jp_k\exp(x_k-x_j).
\end{equation}
The canonical equations of motion read:
\begin{equation}\label{eq x}
\dot{x}_k=\sum_{j<k}p_j\exp(x_k-x_j) +p_k+\sum_{j>k}p_j\exp(x_j-x_k),
\end{equation}
\begin{equation}\label{eq p}
\frac{\dot{p}_k}{p_k}=-\sum_{j<k}p_j\exp(x_k-x_j)+\sum_{j>k}p_j\exp(x_j-x_k).
\end{equation}

One can consider (\ref{eq x}) as a linear system for the impulses $p_k$.
Remarkably enough, the matrix of this linear system has a tridiagonal inverse,
so that $p_k$ depends only on $x_j$, $\dot{x}_j$ with $j=k-1,k,k+1$:
\[
p_k=\dot{x}_k\frac{1-\exp(2(x_{k+1}-x_{k-1}))}
{[1-\exp(2(x_{k+1}-x_k))][1-\exp(2(x_k-x_{k-1}))]}
\]
\begin{equation}\label{imp}
-\dot{x}_{k+1}\frac{\exp(x_{k+1}-x_k)}{1-\exp(2(x_{k+1}-x_k))}-
\dot{x}_{k-1}\frac{\exp(x_k-x_{k-1})}{1-\exp(2(x_k-x_{k-1}))}.
\end{equation}
For $k=1$ and $k=N$ little modifications are necessary, which can be taken
into account by imposing the boundary conditions
\begin{equation}\label{Boundary}
x_0=\infty,\quad x_{N+1}=-\infty.
\end{equation}
The formula (\ref{imp}) implies that the Lagrangian function for the
peakons lattice is local (with only nearest--neighbour contributions):
\[
{\cal L}(x,\dot{x})=
\frac{1}{2}\sum_{k=1}^N\dot{x}_k^2\frac{1-\exp(2(x_{k+1}-x_{k-1}))}
{[1-\exp(2(x_{k+1}-x_k))][1-\exp(2(x_k-x_{k-1}))]}
\]
\begin{equation}\label{Lagr}
-\sum_{k=1}^{N-1}\dot{x}_k\dot{x}_{k+1}\frac{\exp(x_{k+1}-x_k)}
{1-\exp(2(x_{k+1}-x_k))},
\end{equation}

The Lax pair formulation with the matrices from $gl(N)$ for the system
(\ref{eq x})--(\ref{eq p}) was also given in \cite{CH}:
\begin{equation}\label{Lax pair}
\dot{T}=[T,A]
\end{equation}
with the Lax matrix
\begin{equation}\label{Lax matrix}
T(x,p)=\sum_{k=1}^Np_kE_{kk}+
\sum_{j<k}\sqrt{p_kp_j}\exp\left(\frac{1}{2}(x_k-x_j)\right)(E_{kj}+E_{jk}),
\end{equation}
and
\begin{equation}\label{A matrix}
A(x,p)=\frac{1}{2}
\sum_{j<k}\sqrt{p_kp_j}\exp\left(\frac{1}{2}(x_k-x_j)\right)(E_{kj}-E_{jk}).
\end{equation}

An $r$--matrix interpretation of this result was given in \cite{RB}. To
formulate it, let $\langle\cdot,\cdot\rangle$ denote the standard
scalar product in $gl(N)$: $\langle X,Y\rangle={\rm tr}(XY)$, and let the
gradient $\nabla\varphi(T)$ of a smooth function $\varphi$ on $gl(N)$ be
defined by
\[
\langle\nabla\varphi(T),X\rangle=
\left.\frac{d}{d\varepsilon}\varphi(T+\varepsilon X)\right|_{\varepsilon=0}
\quad\forall X \in gl(N).
\]
Let $R(X)$ be a linear operator on $gl(N)$, defined as
\[
{\rm strictly\;lower\;triangular\;part}\;(X)-
{\rm strictly\;upper\;triangular\;part}\;(X),
\]
This operator satisfies the classical modified Yang--Baxter equation,
so that the following Poisson bracket on $gl(N)$ is defined:
\[
\{\varphi,\psi\}(T)=
\frac{1}{2}\Big\langle T, [R(\nabla\varphi(T)),\nabla\psi(T)]+
[\nabla\varphi(T),R(\nabla\psi(T))] \Big\rangle.
\]
Now one of the main results of \cite{RB} states that {\it the set of the
matrices $T(x,p)$ from (\ref{Lax matrix}) forms a Poisson submanifold in
$gl(N)$ equipped with the above Poisson bracket}. Hamiltonian equations
generated in this bracket by a conjugation invariant Hamiltonian function
$\varphi(T)$ read:
\begin{equation}\label{Ham eq}
\dot{T}=\Big[T,\,\frac{1}{2}R(\nabla\varphi(T))\Big]=
\Big[T,\pi_{\pm}(\nabla\varphi(T))\Big],
\end{equation}
where
\[
\pi_+=\frac{1}{2}(R+I),\quad \pi_-=\frac{1}{2}(R-I),\quad {\rm so\;that}\quad
\pi_+-\pi_-=I.
\]
This explains the formulas (\ref{Lax pair})--(\ref{A matrix}), since the
Hamiltonian $H(x,p)$ of the peakons lattice is nothing else then
$\varphi(T)=\frac{1}{2}{\rm tr}(T^2)$, so that $\nabla\varphi(T)=T$.

\setcounter{equation}{0}
\section{Discrete--time peakons lattice}
A general recipe for obtaining  integrable time discretizations for equations
of the form (\ref{Ham eq}) was given in \cite{New discr Toda},\cite{Discr RTL}.
Discrete time systems obtained by this recipe share with the underlying
continuous--time ones the Lax matrix, and hence the integrals of motion
and the integrability property. The formula reads:
\begin{equation}\label{Gen discr eq}
T(t+h)=\Pi_{\pm}^{-1}\Big(I+h\nabla\varphi(T(t))\Big)T(t)
\Pi_{\pm}\Big(I+h\nabla\varphi(T(t))\Big)
\end{equation}
Here the factorization $Y=\Pi_+(Y)\Pi_-^{-1}(Y)$ in the group $GL(N)$
corresponds canonically to the additive decomposition $X=\pi_+(X)-\pi_-(X)$
in its Lie algebra $gl(N)$, and in our case is characterized by the following
conditions: $\Pi_+(Y)$ ($\Pi_-(Y)$) is a nondegenerate lower triangular (resp.
upper triangular) matrix, and the diagonals of these matrices are mutually
inverse.

The aim of the present Letter is to apply this recipe to the peakons
lattice, i.e. to put in the equation (\ref{Gen discr eq}) the Lax
matrix (\ref{Lax matrix}) and $\nabla\varphi(T)=T$.

{\bf Theorem 1.} {\it  The equation
\begin{equation}\label{Discr Lax pair}
T(t+h)=\Pi_+^{-1}\Big(I+hT(t)\Big)T(t)\Pi_+\Big(I+hT(t)\Big)
\end{equation}
serves as a Lax representation for the following map
\footnote{We use the tilde to denote the time $h$ shift, so that
$\wix_k=x_k(t+h)$, if $x_k=x_k(t)$.}
in the space ${\Bbb R}^{2N}\{x,p\}$:
\begin{equation}\label{Discr eq x}
\exp(\wix_k-x_k)=\frac{1}{\beta_k}+hp_k+h\sum_{j>k}p_j\exp(x_j-x_k),
\end{equation}
\begin{equation}\label{Discr eq p}
\frac{\wip_k}{p_k}=\beta_k+\frac{h\beta_k^2}{1+hp_k\beta_k}\;
\sum_{j>k}p_j\exp(x_j-x_k).
\end{equation}
Here the quantities $\beta_k$ are defined recurrently by the relation
\begin{equation}\label{Beta}
\beta_{k+1}=1-\exp(x_{k+1}-x_k)+\frac{\exp(x_{k+1}-x_k)}{hp_k+
\displaystyle\frac{1}{\beta_k}},
\end{equation}
with an initial condition}
\[
\beta_1=1.
\]

{\bf Remark 1.} A simple induction shows that the finite continued
fractions $\beta_k$ have by $h\to 0$ the following asymptotics:
\begin{equation}\label{Asymp}
\beta_k=1-h\sum_{j<k}p_j\exp(x_k-x_j)+O(h^2).
\end{equation}
Hence the discrete time equations (\ref{Discr eq x}), (\ref{Discr eq p})
in fact approximate the continuous time ones (\ref{eq x}), (\ref{eq p}).

{\bf Remark 2.} Due to the general theory (cf. \cite{New discr Toda},
\cite{Discr RTL}), the initial value problem for the discrete time peakons
lattice can be solved in terms of the Cholesky factorization of the matrix
$\left(I+hT(0)\right)^n$, and the interpolating Hamiltonian for this
system is given by ${\rm tr}(\Phi(T))$, where
\[
\Phi(\xi)=h^{-1}\int_0^{\xi}\log(1+h\eta)d\eta=\frac{1}{2}\xi^2+O(h).
\]

{\bf Proof of the Theorem 1.} Due to the symmetry of the matrix $T$, the
$\Pi_+\Pi_-^{-1}$  factorization for $I+hT$ takes the form of the so--called
Cholesky factorization:
\begin{equation}\label{Chol}
I+hT=LL^T,
\end{equation}
where
\begin{equation}\label{L matrix}
L=\Pi_+(I+hT)=\sum_{k\ge j}l_{kj}E_{kj}
\end{equation}
is the (uniquely defined) lower triangular matrix with positive diagonal
entries. With the help of (\ref{Chol}), (\ref{L matrix}) the equation
(\ref{Discr Lax pair}) can be rewritten as
\begin{equation}\label{shifted Chol}
I+h\widetilde{T}=L^TL.
\end{equation}
It turns out that the special structure of the matrix $T$ (\ref{Lax matrix})
allows for an almost explicit computation of the factor $L$.

{\bf Lemma.}

{\rm i)} {\it For} $k>j$
\begin{equation}\label{lkj}
l_{kj}=\sqrt{\frac{p_k}{p_j}}\exp\left(\frac{1}{2}(x_k-x_j)\right)\left(l_{jj}-
\frac{1}{l_{jj}}\right);
\end{equation}

{\rm ii)}
\begin{equation}\label{lkk recur}
l_{k+1,k+1}^2=1+hp_{k+1}-\exp(x_{k+1}-x_k)\frac{p_{k+1}}{p_k}
\left(\frac{1}{l_{kk}^2}-1+hp_k\right).
\end{equation}

The proof of this lemma is relegated to the end of this section. The Theorem 1
simply follows from this lemma. Indeed, setting in
(\ref{lkk recur})
\begin{equation}\label{Diag}
l_{kk}^2=1+hp_k\beta_k,
\end{equation}
we immediately arrive at the recurrent equation (\ref{Beta}) for $\beta_k$.
The following relation, equivalent to (\ref{Diag}), appears to be also useful:
\begin{equation}\label{Diag 1}
\left(l_{kk}-\frac{1}{l_{kk}}\right)^2=
\frac{h^2p_k^2\beta_k^2}{1+hp_k\beta_k}.
\end{equation}

Substituting now (\ref{Lax matrix}) into (\ref{shifted Chol}), we get:
\begin{equation}\label{Tilded elem}
1+h\wip_k=\sum_{i\ge k}l_{ik}^2;\quad
h\sqrt{\wip_k\wip_j}\exp\Big(\frac{1}{2}(\wix_k-\wix_j)\Big)
=\sum_{i\ge k}l_{ik}l_{ij}\;\;{\rm for}\;\;k>j.
\end{equation}
With the help of (\ref{lkj}) the first equation in (\ref{Tilded elem})
may be re--written as
\begin{equation}\label{Aux discr eq}
h\wip_k=l_{kk}^2-1+\left(l_{kk}-\frac{1}{l_{kk}}\right)^2
\sum_{i>k}\frac{p_i}{p_k}\exp(x_i-x_k).
\end{equation}
Now (\ref{Discr eq p}) follows immediately, upon application of (\ref{Diag}),
(\ref{Diag 1}).

The right--hand side of the second equation in (\ref{Tilded elem})  may be
brought with the help of (\ref{lkj}) into the form
\[
\sqrt{\frac{p_k}{p_j}}\exp\Big(\frac{1}{2}(x_k-x_j)\Big)
\left(l_{jj}-\frac{1}{l_{jj}}\right)
\left[l_{kk}+\left(l_{kk}-\frac{1}{l_{kk}}\right)
\sum_{i>k}\frac{p_i}{p_k}\exp(x_i-x_k)\right].
\]
Because of (\ref{Aux discr eq}) this is equal to
\[
\sqrt{\frac{p_k}{p_j}}\exp\Big(\frac{1}{2}(x_k-x_j)\Big)\;
\frac{l_{jj}-\displaystyle\frac{1}{l_{jj}}}
{l_{kk}-\displaystyle\frac{1}{l_{kk}}}\;h\wip_k.
\]
Hence we arrive at the conclusion that the quantity
\[
\frac{\exp(\wix_k-x_k)}
{\wip_kp_k}\left(l_{kk}-\displaystyle\frac{1}{l_{kk}}\right)^2
\]
does not depend on $k$. Setting it equal to $h^2$, we get upon use of
(\ref{Diag 1}):
\begin{equation}\label{Altern discr eq x}
\exp(\wix_k-x_k)=\frac{\wip_k}{p_k}\;\frac{1+hp_k\beta_k}{\beta_k^2}.
\end{equation}
This, together with (\ref{Discr eq p}) is equivalent to
(\ref{Discr eq x}). The Theorem 1 is proved.

{\bf Proof of the lemma.}
The statement i) may be proved by induction in $j$.
Suppose that it holds for $l_{ki}$ with $k>i$ for all $i<j$.
{}From the definition of the matrix $L$ (\ref{Chol}) we have:
\begin{equation}\label{A1}
\sum_{i\le j}l_{ji}^2=1+hp_j;\quad \sum_{i\le j}l_{ki}l_{ji}=
h\sqrt{p_kp_j}\exp\left(\frac{1}{2}(x_k-x_j)\right)\;\;{\rm for}\;\;k>j.
\end{equation}
Using the base of induction, i.e. the expressions (\ref{lkj})  for $l_{ki}$,
$l_{ji}$ with $i<j$, we deduce from the first equation in (\ref{A1}):
\begin{equation}\label{A2}
l_{jj}^2=1+p_j
\bigg[h-\sum_{i<j}\frac{\exp(x_j-x_i)}{p_i}\left(l_{ii}-\frac{1}{l_{ii}}
\right)^2\bigg].
\end{equation}
and from the second one:
\[
l_{kj}l_{jj}=\sqrt{p_kp_j}\exp\left(\frac{1}{2}(x_k-x_j)\right)
\bigg[h-\sum_{i<j}\frac{\exp(x_j-x_i)}{p_i}\left(l_{ii}-\frac{1}{l_{ii}}
\right)^2\bigg].
\]
Comparing the right--hand side of the last equation with (\ref{A2}), we obtain:
\[
l_{kj}=\frac{1}{l_{jj}}\;\sqrt{p_kp_j}\exp\left(\frac{1}{2}(x_k-x_j)\right)\;
\frac{(l_{jj}^2-1)}{p_j}.
\]
This is identical with (\ref{lkj}), which is hence proved for all $k>j$.

The statement ii) follows now from the formula (\ref{A2}):
\[
l_{k+1,k+1}^2-1-hp_{k+1}=
-p_{k+1}\sum_{i\le k}\frac{\exp(x_{k+1}-x_i)}{p_i}
\left(l_{ii}-\frac{1}{l_{ii}}\right)^2=
\]
\[
=-p_{k+1}\exp(x_{k+1}-x_k)\left[\frac{1}{p_k}
\left(l_{kk}-\frac{1}{l_{kk}}\right)^2+
\sum_{i<k}\frac{\exp(x_k-x_i)}{p_i}
\left(l_{ii}-\frac{1}{l_{ii}}\right)^2\right]=
\]
(using again (\ref{A2})  to express the last sum)
\[
=-\frac{p_{k+1}}{p_k}\exp(x_{k+1}-x_k)\left[
\left(l_{kk}-\frac{1}{l_{kk}}\right)^2
-(l_{kk}^2-1-hp_k)\right],
\]
and (\ref{lkk recur}) follows. The lemma is proved.

\setcounter{equation}{0}
\section{Lagrangian formulation}
Recall that in the continuous time case, considering the evolution
equation (\ref{eq x}) as a linear system for $p_k$, we got a ''tridiagonal''
expression (\ref{imp}). In the discrete time case, the corresponding evolution
equation (\ref{Discr eq x}) is highly nonlinear, due to the continued fraction
expression (\ref{Beta}) for $\beta_k$. Nevertheless, it turns out to be
possible to obtain ''tridiagonal'' expressions for the impulses $p_k$ in the
dicrete time case, also.

In order to formulate the corresponding result, it will be convenient to
introduce following short--hand notations:
\begin{equation}\label{A and B}
A_k=\frac{1}{2}\Big(\exp(-x_{k+1})-\exp(-x_k)\Big),\quad
B_k=\frac{1}{2}\Big(\exp(\wix_k)-\exp(\wix_{k+1})\Big).
\end{equation}

{\bf Theorem 2.} {\it There hold following relations:}
\[
hp_k=\frac{1}{2}\Big(\exp(\wix_{k-1}-x_k)-\exp(\wix_{k+1}-x_k)\Big)
\]
\begin{equation}\label{p thru x}
+\exp(-x_k)B_k\sqrt{1+\frac{1}{A_kB_k}}
-\exp(-x_k)B_{k-1}\sqrt{1+\frac{1}{A_{k-1}B_{k-1}}}\;;
\end{equation}
\[
\]
\[
h\wip_k=\frac{1}{2}\Big(\exp(\wix_k-x_{k+1})-\exp(\wix_k-x_{k-1})\Big)
\]
\begin{equation}\label{wip thru x}
-\exp(\wix_k)A_k\sqrt{1+\frac{1}{A_kB_k}}
+\exp(\wix_k)A_{k-1}\sqrt{1+\frac{1}{A_{k-1}B_{k-1}}}\;.
\end{equation}

This Theorem will be proved at the end of this section. Several remarks are
now in turn.

{\bf Remark 3.}
For $k=1$ and $k=N$ minor correction in the formulas (\ref{p thru x}),
(\ref{wip thru x}) are necessary, due to the boundary conditions
(\ref{Boundary}). The corresponding formulas read:
\begin{eqnarray*}
hp_1 & = & -1+\frac{1}{2}\Big(\exp(\wix_1-x_1)-\exp(\wix_2-x_1)\Big)+
\exp(-x_1)B_1\sqrt{1+\frac{1}{A_1B_1}}\;;\\ \\
h\wip_1 & = & \frac{1}{2}\Big(\exp(\wix_1-x_2)+\exp(\wix_1-x_1)\Big)-
\exp(\wix_1)A_1\sqrt{1+\frac{1}{A_1B_1}}\;;\\ \\
hp_N & = & \frac{1}{2}\Big(\exp(\wix_{N-1}-x_N)+\exp(\wix_N-x_N)\Big)
-\exp(-x_N)B_{N-1}\sqrt{1+\frac{1}{A_{N-1}B_{N-1}}}\;;\\ \\
h\wip_N & = & -1+\frac{1}{2}\Big(\exp(\wix_N-x_N)
-\exp(\wix_N-x_{N-1})\Big)
+\exp(\wix_N)A_{N-1}\sqrt{1+\frac{1}{A_{N-1}B_{N-1}}}\;.
\end{eqnarray*}

{\bf Remark 4.}
The expressions (\ref{p thru x}), (\ref{wip thru x}) serve as
finite--difference approximations to (\ref{imp}). However, as opposed to the
analogous statements in \cite{New discr Toda}, \cite{Discr RTL}, this is not
immediately obvious, but rather requires for some (straightforward)
calculations.

Now we are in a position to give a Lagrangian formulation of the discrete
time peakons lattice, i.e. to represent it in the form
\begin{equation}\label{Discr Lagr eq}
\partial\Big(\Lambda(x(t+h),x(t))+\Lambda(x(t),x(t-h))\Big)/\partial x_k(t)=0.
\end{equation}
The key point is the representation of the momenta in terms of the Lagrangian
function \cite{BRST}:
\begin{equation}\label{Momenta}
p_k=-\partial\Lambda(\wix,x)/\partial x_k,\quad
\widetilde{p}_k=\partial\Lambda(\wix,x)/\partial \wix_k.
\end{equation}
Identifying these expressions with (\ref{p thru x}), (\ref{wip thru x}),
we see that the following statement holds.

{\bf Theorem 3.} {\it The discrete time peakons lattice is a Lagrangian
system (\ref{Discr Lagr eq}) with the Lagrangian function
\[
\Lambda(\wix,x)=x_1+\frac{1}{2}\exp(\wix_1-x_1)+
\frac{1}{2}\exp(\wix_N-x_N)-\wix_N
\]
\begin{equation}\label{Lagrangian}
+\frac{1}{2}\sum_{k=1}^{N-1}\Big(\exp(\wix_k-x_{k+1})-\exp(\wix_{k+1}-x_k)\Big)
-\sum_{k=1}^{N-1}\Psi(A_kB_k),
\end{equation}
where}
\[
\Psi(\xi)=2\int_0^{\xi}\sqrt{1+\frac{1}{\eta}}d\eta.
\]

{\bf Remark 5.}
The function (\ref{Lagrangian}) serves as a rather non--trivial approximation
to (\ref{Lagr}).
This can be seen after some straightforward, though tediuos calculations.

{\bf Proof of the Theorem 2.} Subtract from the equation (\ref{Discr eq x})
the analogous equation for the subscript $k+1$ multiplied by
$\exp(x_{k+1}-x_k)$:
\[
\left(\frac{1}{\beta_k}+hp_k\right)-\frac{1}{\beta_{k+1}}\exp(x_{k+1}-x_k)=
\exp(\wix_k-x_k)-\exp(\wix_{k+1}-x_k).
\]
Using the recurrence (\ref{Beta}) to express $\beta_{k+1}$, we end up after
some simple calculations with the following quadratic equation:
\[
\left(\frac{1}{\beta_k}+hp_k\right)^2-
2\exp(-x_k)B_k\left(\frac{1}{\beta_k}+hp_k\right)-
\exp(-2x_k)\frac{B_k}{A_k}=0.
\]
Its (positive) solution:
\begin{equation}\label{1beta plus p}
\frac{1}{\beta_k}+hp_k=\exp(-x_k)B_k\left(\sqrt{1+\frac{1}{A_kB_k}}+1\right).
\end{equation}
In order to exclude $\beta_k$ from this expression, we use (\ref{Beta}) in
the form
\[
\beta_k=\exp(x_k)\left(2A_{k-1}+
\frac{\exp(-x_{k-1})}{\displaystyle\frac{1}{\beta_{k-1}}+hp_{k-1}}\right).
\]
Using in the right--hand side (\ref{1beta plus p}), we get
\begin{equation}\label{beta}
\beta_k=\exp(x_k)A_{k-1}\left(\sqrt{1+\frac{1}{A_{k-1}B_{k-1}}}+1\right),
\end{equation}
or
\begin{equation}\label{1beta}
\frac{1}{\beta_k}=\exp(-x_k)B_{k-1}
\left(\sqrt{1+\frac{1}{A_{k-1}B_{k-1}}}-1\right).
\end{equation}
Now (\ref{p thru x}) follows directly from (\ref{1beta plus p}) and
(\ref{1beta}).

Finally, to prove (\ref{wip thru x}) we use (\ref{Altern discr eq x}),
re--written in the form
\[
h\wip_k=hp_k\exp(\wix_k+x_k)\frac{\beta_k\exp(-x_k)}
{\left(\displaystyle\frac{1}{\beta_k}+hp_k\right)\exp(x_k)}.
\]
Putting (\ref{p thru x}), (\ref{1beta plus p}), (\ref{beta}) into this formula,
we arrive after some manipulations at (\ref{wip thru x}). The Theorem 2
is proved.

\section{Conclusion}
A new application of a general scheme for producing integrable discretizations
for integrable Hamiltonian flows is described in the present Letter.
Advantages of this approach are rather obviuos: it is, in principle, applicable
in a standartized way to every system admitting an $r$--matrix formulation,
at least with a constant $r$--matrix satisfying the modified Yang--Baxter
equation. We shall demonstrate elsewhere that the discrete time systems from
\cite{Discr CM}, \cite{Discr RS} with dynamical $r$--matrices may be also
included into this framework. We hope also to report on numerous other
applications of this approach in the future.

The drawback of this scheme is also obvious to any expert in this field.
Namely, some of the most beautiful discretizations do not live on the same
$r$--matrix orbits as their continuous time counterparts \cite{Ves et al},
\cite{Discr Toda}, \cite{Ragn}, \cite{Discr Garnier}, and there seems to
exist no way of {\it a priori} identifying the correct orbit for nice
discretizations. However, we hope that continuing to collect examples will
someday bring some light to this intriguing problem.

The research of the author is financially supported by the DFG (Deutsche
Forschungsgemeinshaft).

\end{document}